\documentstyle[12pt,epsf,epsfig]{article}
\newcount\timecount
\newcount\hours \newcount\minutes  \newcount\temp \newcount\pmhours

\hours = \time
\divide\hours by 60
\temp = \hours
\multiply\temp by 60
\minutes = \time
\advance\minutes by -\temp
\def\hour{\the\hours}
\def\minute{\ifnum\minutes<10 0\the\minutes
            \else\the\minutes\fi}
\def\clock{
\ifnum\hours=0 12:\minute\ AM
\else\ifnum\hours<12 \hour:\minute\ AM
      \else\ifnum\hours=12 12:\minute\ PM
            \else\ifnum\hours>12
                 \pmhours=\hours
                 \advance\pmhours by -12
                 \the\pmhours:\minute\ PM
                 \fi
            \fi
      \fi
\fi
}

\def\monthname{\relax\ifcase\month 0/\or January\or February\or
   March\or April\or May\or June\or July\or August\or September\or
   October\or November\or December\else\number\month/\fi}

\def\bold#1{\setbox0=\hbox{$#1$}%
     \kern-.025em\copy0\kern-\wd0
     \kern.05em\copy0\kern-\wd0
     \kern-.025em\raise.0433em\box0 }

\def\ga{\mathrel{\raise.3ex\hbox{$>$\kern-.75em\lower1ex\hbox{$\sim$}}}}
\def\la{\mathrel{\raise.3ex\hbox{$<$\kern-.75em\lower1ex\hbox{$\sim$}}}}
\def\gev{{\rm \, Ge\kern-0.125em V}}
\def\tev{{\rm \, Te\kern-0.125em V}}
\def\beq{\begin{equation}}
\def\eeq{\end{equation}}

\def\ss{\scriptscriptstyle}

\def\ohsq{\Omega_{\chi} h^2}

\def\m12{m_{1\!/2}}

\def\tb{\tan\beta}

\begin{document}
\begin{titlepage}
\pagestyle{empty}
\baselineskip=21pt
\rightline{hep-ph/9801445}
\rightline{CERN--TH/98-32}
\rightline{MPI-PhE/98-1}
\rightline{UMN--TH--1619/98}
\rightline{MADPH-98-1032}
\vskip 0.2in
\begin{center}
{\large{\bf
Charginos and Neutralinos in the Light of Radiative Corrections:\\
Sealing the Fate of Higgsino Dark Matter}}
\end{center}
\begin{center}
\vskip 0.1in
{{\bf John Ellis}$^1$, {\bf Toby Falk}$^2$, {\bf Gerardo Ganis}$^3$,
{\bf Keith A.~Olive}$^4$
and {\bf Michael Schmitt}$^5$}\\
\vskip 0.1in
{\it
$^1${TH Division, CERN, Geneva, Switzerland}\\
$^2${Department of Physics, University of Wisconsin, Madison, WI~53706,
Wisconsin}\\
$^3${Max-Planck-Institut f\"ur Physik, Munich, Germany}\\
$^4${School of Physics and Astronomy,
University of Minnesota, Minneapolis, MN 55455, USA}\\
$^5${PPE Division, CERN, Geneva, Switzerland;\\
Address after Feb. 1st, 1998: Physics Department, Harvard University,
Cambridge, MA~02138, USA}\\}
\vskip 0.2in
{\bf Abstract}
\end{center}
\baselineskip=18pt \noindent
We analyze the LEP constraints from searches for charginos
$\chi^{\pm}$ and neutralinos $\chi_i$, taking into account radiative
corrections to the relations between their masses and the underlying
Higgs-mixing and gaugino-mass parameters $\mu, m_{1/2}$ and the
trilinear mass parameter $A_t$.  Whilst radiative corrections do not
alter the excluded domain in $m_{\chi^{\pm}}$ as a function of
$m_{\chi^{\pm}} - m_{\chi}$, its mapping into the $\mu, m_{1/2}$
plane is altered.  We update our previous lower limits on the mass of
gaugino dark matter and on tan$\beta$, the ratio of Higgs vacuum
expectation values, in the light of the latest LEP data and these
radiative corrections.  We also discuss the viability of
Higgsino dark matter, incorporating co-annihilation effects into the
calculation of the Higgsino relic abundance.  We find that Higgsino
dark matter is viable for only a very limited range of $\mu$ and
$m_{1/2}$, which will be explored completely by upcoming LEP runs.
\vfill
\vskip 0.15in
\leftline{CERN--TH/98-32}
\leftline{January 1998}
\end{titlepage}
\baselineskip=18pt
\section{Introduction}

Considerable experimental effort is currently being devoted at LEP and
elsewhere to the search for the charginos $\chi^{\pm}$ and neutralinos
$\chi_i \; ; \; i = 1,..,4$ of the minimal supersymmetric extension of
the Standard Model (MSSM)~\cite{latest,contr,LEPC,Janot}, and to
understanding the
constraints these and other particle searches impose on the parameter
space of the MSSM~\cite{EFOSI,EFOSII}. One of the generic
possibilities for MSSM~\cite{MSSM} phenomenology is that supersymmetry
breaking is communicated to the observable sparticles via
gravitational interactions, in which case the lightest supersymmetric
particle is presumably the lightest neutralino $\chi_1$ (hereafter $\chi$), which would
be stable if $R$ parity is conserved.  The lightest neutralino is not
directly visible to accelerator experiments in this type of
$R$-conserving supergravity scenario, and would be a good
candidate~\cite{hg,EHNOS} for the Cold Dark Matter advocated by many
astrophysicists and theorists of cosmological structure
formation~\cite{CDM}, since it might well have $0.1 \le \ohsq \le
0.3$, where $\Omega_{\chi}$ is the relic density of the lightest
neutralino $\chi$ in units of the critical density, and $h$ is the
present Hubble expansion rate in units of 100 km/s/Mpc. Within this
framework, which we adopt in this paper, detailed modelling of the
relations between different searches is often necessary in order to
close certain loopholes in the MSSM parameter space, in particular if
one wishes to establish robust lower limits on the mass of the
lightest neutralino $\chi$~\cite{ALEPH,Janot}.  Such lower limits on
$m_{\chi}$ are of particular relevance to the ongoing direct and
indirect non-accelerator searches for cosmological relic
neutralinos~\cite{CDM}.

The strength of the lower limit
obtained depends~\cite{EFOSI,EFOSII} on various theoretical assumptions,
such as the
degree of universality among input supersymmetry-breaking
parameters, and whether one requires that the
relic density of neutralinos lie within the range favoured by
astrophysical and cosmological considerations~\cite{CDM}.
The successive upgrades of the LEP~2 centre-of-mass 
energy~\cite{latest,contr,LEPC} have
enabled this lower limit to be strengthened
progressively, and
it currently stands at $m_{\chi} \ga 40$ GeV if the strongest
versions of these theoretical and astrophysical assumptions are
imposed~\cite{EFOSII}.

Although radiative corrections to MSSM Higgs scalar
masses~\cite{MSSMHiggs} have
been included in the analysis of LEP data~\cite{Janot}, 
and play an important r\^ole,
radiative corrections to the masses of the
charginos and neutralinos have not so far been included.
The one-loop corrections
to the relations between the physical masses and underlying
MSSM parameters such as the gaugino masses $M_a: a = 1,2,3$ and the
Higgs-mixing
parameter $\mu$ are by now well known~\cite{LTT,PP,PBMZ} and
non-negligible.~\footnote{Radiative corrections to the $\chi^+ \chi^-$
production cross section are also available~\cite{DKR}, but appear
less crucial
at present, as we discuss below.} In
particular, they are known to modify considerably the tree-level
difference between the lightest chargino and neutralino masses,
$m_{\chi^{\pm}} - m_{\chi}$, which is one of the most important
parameters controlling the efficiencies of experimental searches,
especially in the ``Higgsino'' region of small $\mu$ and 
relatively large $M_a$~\cite{GP,DNRY}.
More generally, the radiative corrections to the chargino and neutralino
masses cause shifts in the inferred exclusion domains in the
$\mu, M_2$ plane that are not negligible compared to
those deduced from the improvements in the chargino mass limits obtained
after successive LEP~2 energy upgrades. These changes are
also comparable to the differences between mass limits obtained under
different theoretical assumptions. Therefore, radiative corrections to the
chargino and neutralino mass relations could in principle play a r\^ole
in the delicate interplay of the direct experimental constraints used to
constrain
$m_{\chi}$~\cite{ALEPH,EFOSI}. Clearly the degree of effort put into the
experimental
searches~\cite{latest,contr,LEPC} and their phenomenological
interpretation merits the
inclusion of these radiative corrections in the analysis of the
LEP~2 data.

We find that there are significant shifts in the allowed regions of
the $\mu, M_2$ plane. The effects on
$m_{\chi^{\pm}}$ are particularly important
when it is analyzed in terms of $M_2$, but not when it is
analyzed as a function of $\Delta M \equiv m_{\chi^{\pm}} - 
m_{\chi}$.
We use these radiative corrections in an update of
our previous analysis of LEP constraints in the MSSM parameter space,
and on $m_{\chi}$ and tan$\beta$ in particular, assuming 
universal input soft supersymmetry-breaking masses and a plausible
cosmological relic density. This update also includes
the most recent preliminary limits on chargino and neutralino
production from LEP running at 183~GeV~\cite{latest,contr,LEPC,Janot}, as
well as new limits
on the lightest neutral Higgs mass $m_h$~\cite{latesth,contrh,LEPC,Janot}. 
Relaxing the universal scalar mass assumption, we focus on
the portion of the MSSM parameter space for which the lightest
neutralino is a Higgsino, where the limits are particularly
sensitive to the radiative corrections~\cite{GP,DNRY}. In this
context, we evaluate the possibility of Higgsino dark matter,
incorporating both co-annihilations and radiative corrections in
conjunction with the bounds imposed by LEP. We find only a very
limited range of parameters in which Higgsino dark matter is
viable, which will be explored completely by forthcoming LEP runs.

\section{Implementation of Radiative Corrections in Chargino and
Neutralino Codes for LEP}

As already mentioned, we work in the context of the MSSM in a supergravity 
framework where supersymmetry breaking is communicated to observable
sparticles by gravity, so that the lightest neutralino
$\chi$ is the lightest supersymmetric particle, and we also
assume that
$R$ parity is conserved, so that $\chi$ is stable. We further assume
that the supersymmetry-breaking gaugino mass
parameters are universal at the input supergravity scale: 
$M_a = m_{1/2}$. Similarly, we assume that the 
soft supersymmetry-breaking contributions to the
squark and slepton masses $m_{\tilde q}, m_{\tilde
\ell}$ are universal and equal to $m_0$ at the unification scale, and the
trilinear parameters $A$ are also taken to be
universal. The MSSM can then be parameterized by
$m_{1/2}, m_0, A$, $\mu$, the pseudoscalar Higgs mass
$m_A$, and $\tan \beta$.
Here we project the experimental and cosmological bounds onto the
$\mu, \, M_2$ plane for fixed $m_0, A$, $m_A$, and $\tan \beta$. 

We note that only the trilinear coupling $A_t$, associated with the
top-quark Yukawa coupling, is relevant for the phenomenology discussed
here~\footnote{We do not consider very large values of tan$\beta\gg
  10$.}, and that the possible range of values of this parameter is
constrained because of an infrared quasi-fixed point: $A_t \sim 2
m_{1/2} \sim 2 m_{\tilde g} / 3$~\cite{irqfp}, so that this parameter
is inessential.  This quasi-fixed point arises because, if $A_t$ is much
larger than $\m12$, the leading-order running of $A_t$ is given by
$dA_t/dt\approx 6 \lambda_t^2 A_t/8\pi^2$~\cite{MSSM}, so that (for
constant $\lambda_t$) $A_t$ decreases as a power law with scale, with
an exponent that is roughly 1/12 for $\lambda_t \sim 1$.  Thus, if 
the input $A_t\gg
\m12$, $A_t$ is reduced by an order of magnitude in its evolution from
$M_X$ to $M_Z$.  There are additional negative terms in the beta
function for $A_t$ which are proportional to the gaugino masses, 
so that $\m12$ sets the scale for $A_t(M_Z)$.  Consequently,
$A_t(M_Z) \gg \m12$ requires extremely large $A_t(M_X)$. However,
this tends to drive the stop soft mass-squared parameters negative at
$M_Z$, so
large values of $A_t(M_Z)$ are not allowed, at least for the
values of $\tan\beta$ we consider.  Therefore,
we do not explore a range in values for $A_t$, but
concentrate on the fixed-point value.

The input supersymmetry-breaking parameters are evolved down to the
electroweak scale using the two-loop renormalization-group equations
for the
gaugino masses, so that the one-loop
relation $M_a = (\alpha_a / \alpha_{GUT})\times m_{1/2}$ is violated
at the 10\% level. We also evolve the gauge and Yukawa couplings at
the two-loop level, but other parameters are evolved with one-loop
accuracy.  At the tree level, we parameterize chargino and neutralino
eigenstates in terms of $M_2, \mu$ and tan$\beta$. We take a top mass
of 171 $\gev$: the only significant sensitivity of our results to $m_t$
comes in the Higgs mass constraints, which we discuss in Section 5.

This theoretical framework has been adopted in many of
the searches for supersymmetric particles at LEP,
in particular for charginos and neutralinos.
Until now, all the experimental analyses have used
tree-level formulae for the masses and cross sections,
in terms of the MSSM parameters introduced above.
The one-loop calculations of the physical
masses of the charginos and neutralinos~\cite{LTT,PP,PBMZ}, and
more recently the $\chi^+\chi^-$ production cross section~\cite{DKR},
that have become available have been calculated in
the $\overline{DR}$ prescription, which is the supersymmetric
analogue of the conventional $\overline{MS}$ prescription.
The one-loop mass formulae have now been
incorporated~\cite{FG} in one of the codes available for LEP experimental
analyses, and we discuss here a first exploration
of their implications for sparticle mass limits. The radiative
corrections to the cross section are less important for this purpose:
they are generally $\la 15$ \% or so~\cite{DKR}, whereas the
experimental upper
limits on $\sigma_{+-} \equiv \sigma(e^+e^-\rightarrow \chi^+ \chi^-)$ are
generally
far below the theoretical predictions, except very close to the
production threshold: $E_{CM} \sim 2 m_{\chi^{\pm}}$. Since
$\sigma_{+-}$ varies very rapidly with $m_{\chi^{\pm}}$ in this region,
the change the theoretical prediction for $\sigma_{+-}$ has little
effect on the chargino mass limit obtained.

\section{Exploration of Consequences of the Radiative Corrections}

\begin{figure}[thbp]
\begin{center}
 \vspace*{-1.5in}
\hspace*{-0.0in}
\epsfig{file=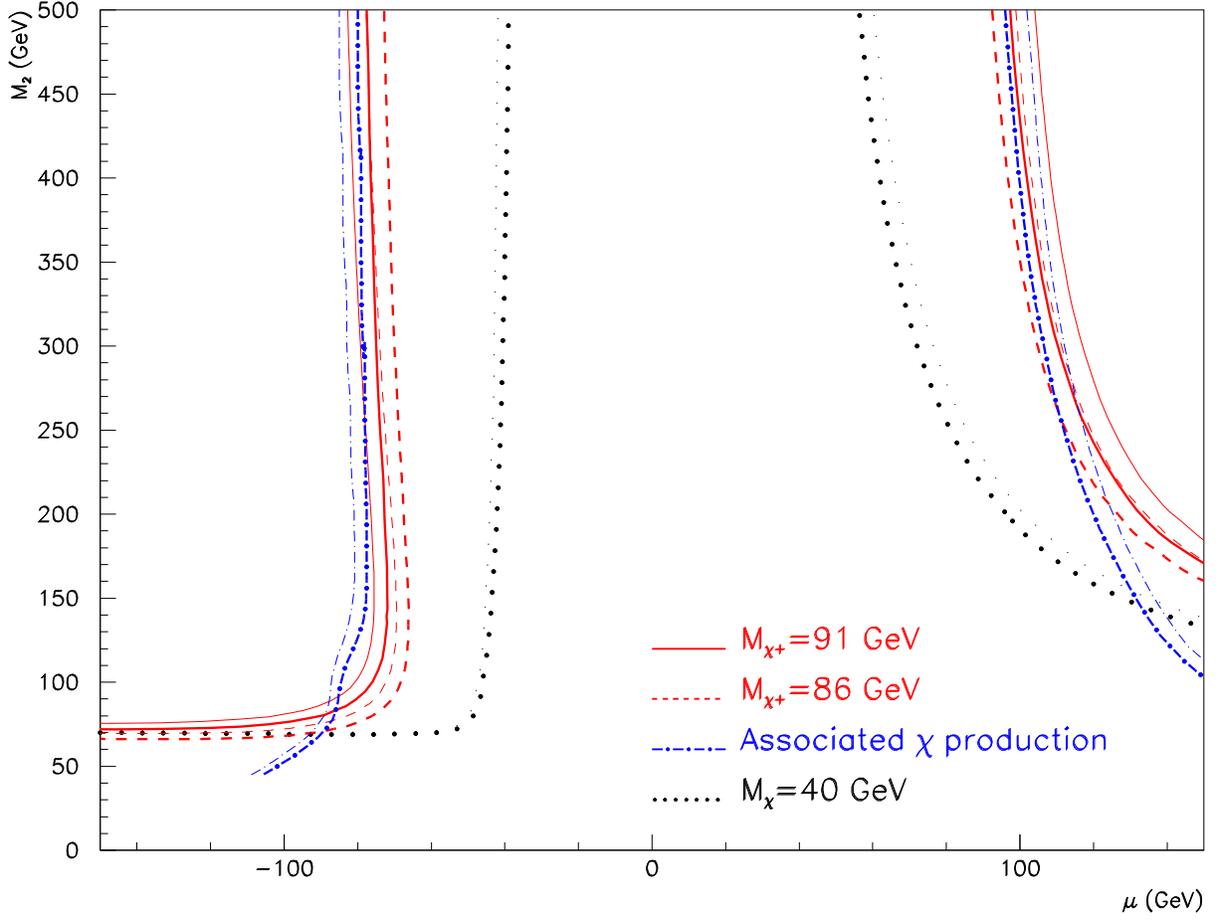,height=5.5in}
\end{center}
\caption{{\it The effects of one-loop
radiative corrections in the $\mu,M_2$ plane, calculated for tan$\beta =
2$, $m_0 = 200$~GeV, $m_A = 1$~TeV and $A_t = 2 m_{1/2}$. The thick lines
are one-loop-corrected
contours corresponding to fixed values of chargino and neutralino
masses, and the thin lines are tree-level contours. The continuous
lines are for $m_{\chi^{\pm}} = 91\gev$, the dashed lines for
$m_{\chi^{\pm}} = 86\gev$, the dash-dotted lines for
the bound on associated neutralino production, and
the dotted lines for $m_{\chi} = 40\gev$.}}
{ \label{fig:muM2} }
\end{figure}

Fig.~\ref{fig:muM2} displays the effects of these one-loop radiative
corrections in the $\mu,M_2$ plane. The thick lines are one-loop-corrected
contours corresponding to fixed values of chargino and neutralino
masses, and the thin lines are tree-level contours. The continuous
lines are for $m_{\chi^{\pm}} = 91\gev$, the dashed lines for
$m_{\chi^{\pm}} = 86\gev$, corresponding 
respectively to the kinematic limits on
$m_{\chi^{\pm}}$ from $e^+e^-
\rightarrow \chi^+ \chi^-$ at the just completed 183~GeV run of LEP
and its
previous run around 172~GeV. The dash-dotted lines show the contour of 
$m_{\chi} = 40$~GeV, corresponding to our 
previous lower limit on $m_{\chi}$ when
several theoretical and cosmological constraints are applied
\cite{EFOSII}.

In general, each of the four neutralinos $\chi_i$ is 
characterized by the four components
$\alpha_i, \beta_i, \gamma_i$, and $\delta_i$ of its
eigenvector in the ${\tilde W}, {\tilde B}, {\tilde H}_1,
{\tilde H}_2$ basis. In this notation, the gaugino fraction of a
particular
neutralino $\chi_i$ is $\sqrt{{\alpha_i}^2 + {\beta_i}^2}$, and its Higgsino
fraction is $\sqrt{{\gamma_i}^2 + {\delta_i}^2}$.  
In a later section of this paper, we discuss whether LEP
limits still allow the lightest neutralino to be 
predominantly a Higgsino. With a view to this
subsequent discussion, we recall here that the
$Z^0$ coupling to a pair of neutralinos
$\chi_i$ and $\chi_j$ is proportional to the Higgsino components
of the neutralinos, namely $\gamma_i \gamma_j - \delta_i \delta_j$.
Thus it is particularly easy for LEP to
probe regions of the $\mu,M_2$ plane where the lightest
neutralino is mainly a
Higgsino, by searching for production in association with
a heavier neutralino state which is also mainly a
Higgsino. Accordingly, we also plot in 
the relevant parts of Fig.~\ref{fig:muM2}
dotted lines representing the experimental limit of 0.18~pb on
associated neutralino production $e^+e^- \rightarrow \chi_i \chi_j: i+j
>2$. In the region of interest for this analysis, this
is dominated by $e^+e^- \rightarrow \chi \chi_{2,3,4}$ and corresponds
essentially to $m_{\chi} + m_{\chi'_H} = 182\gev$,
where $\chi'_H$ is the lightest mainly-Higgsino state
among the $\chi_{2,3,4}$.
%

The one-loop radiative corrections in Fig.~\ref{fig:muM2}
are calculated with the following
representative values of the input MSSM parameters: tan$\beta = 2$,
$m_0= 200$~GeV, $m_A = 1$~TeV and $A_t = 2 m_{1/2}$.
We see significant shifts between the 
pairs of corresponding thick and thin lines,
making manifest the significance of the one-loop radiative corrections.
Varying
$m_0$ between 100 and 1000 GeV does not have a noticeable effect on
the lines plotted, and they do not depend
strongly on the precise value chosen for
$m_A$, though they would be significantly different for
values of $A_t$ much larger than the fixed-point value that we favour.
Interestingly, the differences between these pairs of lines are
comparable to the splitting between the lines for $m_{\chi^{\pm}} = 86$
and $91$~GeV, indicating that the radiative corrections have
phenomenological implications for the 
inferred constraints on the input MSSM parameters $\mu, M_2$ that are
comparable to the upgrade in the LEP energy from 172 to
183~GeV.

\begin{figure}
\vspace*{0.3in}
\begin{center}
\hspace*{0.5in}
\epsfig{file=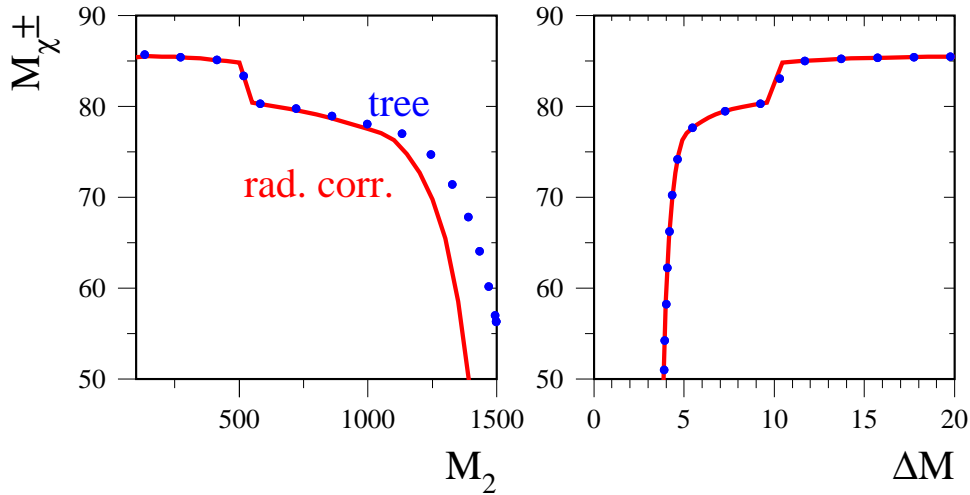,height=2.8in}
\end{center}
\vspace*{-1.4in}
\caption{{\it (a) The experimental limit on $m_{\chi^{\pm}}$ as
function of $M_2$ and as a function of $\Delta M \equiv m_{\chi^{\pm}} - m_{\chi}$,
for fixed $m_0 = 200\gev$  and $\tan\beta = 2$. The value of $\mu$ is
determined by the combination $m_{\chi^{\pm}}$ and $M_2$.}
{\it The drop at large $M_2$ is due to the loss in
experimental efficiency as $\chi^{\pm}$ decays into $\chi$ involve
softer leptons. The dotted line comes from a tree-level analysis, and the
solid line is obtained using an ad hoc parameterization of
the experimental efficiency, in
conjunction with the
radiatively-corrected mass formulae. 
Note that the lines separate significantly
when $M_2$ is large. (b) The
experimental limit on $m_{\chi^{\pm}}$ is now plotted as a function of
$\Delta M \equiv m_{\chi^{\pm}} - m_{\chi}$. Note that the tree-level
and radiatively-corrected curves are almost coincident.}}

{ \label{fig:mchiM2} }
{ \label{fig:mchiDelM} }
\end{figure}

The LEP exclusions for charginos do not always extend
to the kinematic limits. In particular, the lower limit on
$m_{\chi^{\pm}}$ is
sensitive to $m_0$:  it affects $\sigma_{+-}$ through
the ${\tilde \nu}$ exchange diagram, and when $m_0$ is
such that 
$m_{\chi^{\pm}}-m_{\tilde\nu}<3\gev$,  chargino decays to a sneutrino
and a low-energy lepton reduce the experimental efficiency to essentially
zero.
Moreover, as seen in Fig.~\ref{fig:mchiM2},
there is also a loss of sensitivity for fixed $m_0$ at large $M_2$,
because of the reduction in detection efficiency
that occurs when the difference in mass between the produced
particle and its supersymmetric decay product, e.g., $\Delta 
M \equiv m_{\chi^{\pm}} - m_{\chi}$, is small. This implies that at
large $M_2$, beyond the top of Fig.~\ref{fig:muM2}, the experimental
exclusion in the $\mu,M_2$ plane would contract toward $|\mu| \simeq
45\gev$. It has been observed previously~\cite{GP,DNRY} that $\Delta M$
may be altered significantly by the one-loop
radiative corrections, altering the value of $M_2$ where this contraction
would occur. 
For $\tan\beta$ not too large, the leading contribution to the change in
$\Delta M$, $\delta(\Delta M)$, comes
from heavy quark-squark loops, and for split stop masses it is approximately
given by ~\cite{GP,DNRY}
\begin{equation}
\delta(\Delta M) = {3\over 32 \pi^2} \,\lambda_t^2\, m_t \sin{2\theta_t} \,
\log{\max(m_t^2,m_{\tilde t_2}^2)\over m_{\tilde t_{\ss 1}}^2}\, ,
\label{deltaM}
\end{equation}
where $m_{\tilde t_1}>m_{\tilde t_2}$. Note that 
this vanishes in the limit that there is no stop mixing 
($\sin{2 \theta_t}\rightarrow 0$), and that the sign
of $\delta(\Delta M)$ depends on the sign of $\theta_t$.  
Both of these depend on the
off-diagonal element of the stop mass-squared matrix, and hence on $A_t$.
Hence, the quasi-fixed point for $A_t$ 
is an important constraint on
radiative corrections to $\Delta M$.

Incorporating these radiative corrections
fully in the analysis of experimental data would require the
use in experimental simulations of the full radiative-correction
code~\cite{FG}. In the absence so far of such an analysis, we have used
this code in conjunction with an {\it ad hoc} parameterization of the 
efficiency of the ALEPH chargino search~\cite{latest} as a function of
$\Delta M$ to
explore the possible significance of this effect. We see in
Fig.~\ref{fig:mchiM2}a that the exclusion contour  in the $M_2,
m_{\chi^{\pm}}$ plane is indeed affected significantly. However, as seen
in  Fig.~\ref{fig:mchiDelM}b,  this effect of the radiative corrections
vanishes in the $\Delta M, m_{\chi^{\pm}}$ plane, as expected
from the small change in the chargino composition. This
suggests that a useful rule of thumb may be to expect that
the exclusion domain in this particular plane may be insensitive to
the radiative corrections also in a complete experimental
analysis using~\cite{FG}.

%
\section{Updating Lower Limits on $m_{\chi}$ and tan$\beta$}

We now review the lower limit on $m_{\chi}$ derived previously~\cite{EFOSII}
using data from the LEP runs at energies up to 172 GeV, discussing
two issues: the significance of the radiative corrections and the
impact of the more recent data at 183 GeV. The purely experimental
constraint from LEP 172 coming from chargino production and
associated neutralino production yielded the limit $m_{\chi} > 14
\gev$~\cite{latest}, with the minimum occurring around $\tan \beta = 3$.
Adding the theoretical assumption that all squark masses
are universal at the unification scale, unsuccessful Higgs
searches indirectly strengthened
the bound on $m_{\chi}$ significantly in
the region around $\tan \beta = 2$~\cite{EFOSII}. Furthermore, when scalar
mass
universality was extended to the Higgs sector and the
values of $\mu$ and $m_A$ were fixed 
by requiring the consistency of
electroweak symmetry breaking~\cite{dynewsb}, the bound on
$m_{\chi}$ was improved for all $\tan \beta$. For $\tan \beta \ga
4$, we found $m_{\chi} \ga 30\gev$, and the bound improved to
$m_{\chi} \ga 40\gev$ for $\tan \beta \la 3$~\cite{EFOSII}. 

We have also considered imposing the cosmological
constraint that the neutralino relic density
does not over-close the Universe. Specifically, the combination
$\Omega_{\chi} h^2$ is constrained to be less than 0.3 by the
requirement that the age of the Universe, $t_0$, be greater than about 12
Gyr. These quantities are related through
\begin{equation}
H_0 t_0=\int^1_0 dx (1 - \Omega + \Omega/x)^{-1/2},
\end{equation}
where the scaled Hubble parameter is $h = H_0/(100$km Mpc$^{-1}$
s$^{-1}$), and large relic densities may be incompatible with constraints
on $t_0$.
When the cosmological constraint was combined with the experimental
constraints from Higgs searches and the theoretical assumption
of universal Higgs masses,  we found~\cite{EFOSII} that values of
$\tan \beta \la 1.7$ for
$\mu < 0$ and $\la 1.4$ for $\mu > 0$ were excluded.
This is because the
tree-level Higgs mass is small at low $\tan\beta$, 
so radiative corrections to $m_h$ must be
enhanced by taking a large stop mass.  On the other hand, large
sfermion masses lead to large relic densities for a gaugino-like
neutralino, as expected for universal scalar masses. Thus 
$\Omega_{\chi} h^2 < 0.3$ is not compatible with a heavy enough
Higgs boson.
Finally, when we imposed the astrophysical desire that the
cosmological relic neutralinos play a significant role in
structure formation, which requires that
$\Omega_{\chi} h^2 > 0.1$, we found that the
lower limit on the neutralino mass was strengthened at $\tan \beta \ga 3$
to $m_{\chi} \ga 50\gev$.  

In the interests of simplicity, we re-discuss
here the lower bound only in the case in which all
theoretical and cosmological assumptions are made, in
particular that the input scalar masses are universal and that the relic
cosmological density lies in the range $0.1 \le \ohsq
\le 0.3$. The resulting bound on the LSP mass in this case was 
previously found to be $m_{\chi}
\ga 40$~GeV, with the minimum being achieved in a narrow range of
$2.6\la\tan\beta\la3.0$, as seen in Fig.~1 of~\cite{EFOSII}. In
this case, the lower bound on
$m_{\chi}$ was linked directly to the lower limit $m_{\chi^{\pm}} \ga
86$~GeV from the LEP 172 GeV run and relied on the Higgs mass constraint.
Cosmology did not play a direct role for this range of tan$\beta$, though
these constraints did play an important indirect role in closing off
loopholes
for special values of other MSSM parameters such as $\mu$ and $m_0$
where the limit on
$m_{\chi^{\pm}}$ and hence $m_{\chi}$ could have been smaller,
and provided stronger lower limits on $m_{\chi}$ for both lower and
higher values of tan$\beta$.

\begin{figure}[ht]
\vspace*{-.2in}
\begin{center}
\hspace*{-0.5in}
\epsfig{file=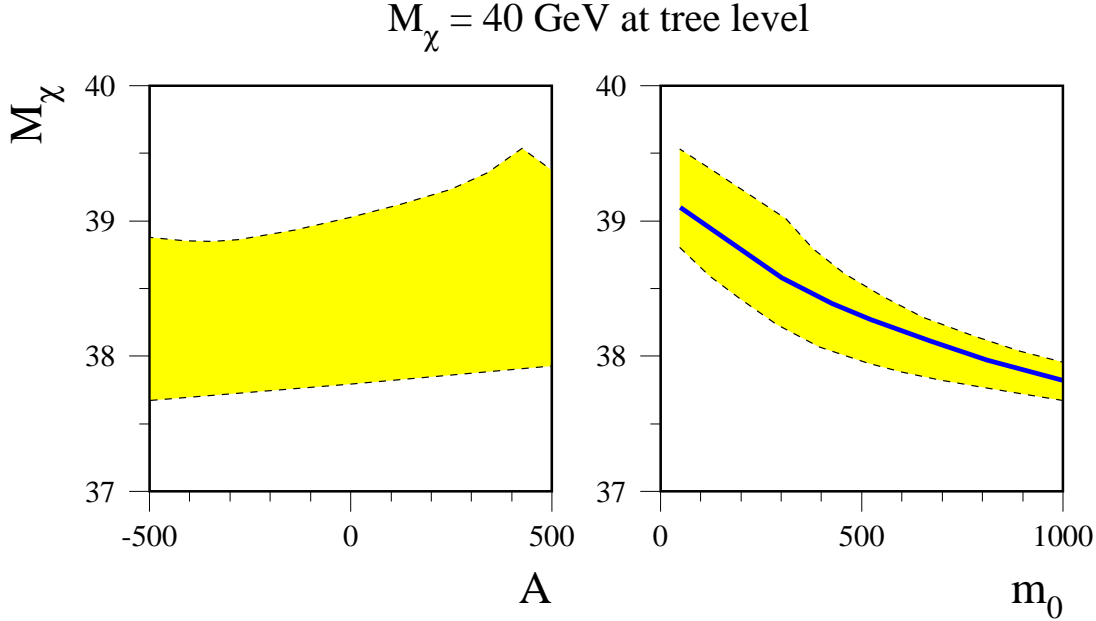,height=3.5in}
\end{center}
\vspace*{-.5in}
 \caption{{\it Plots of the radiative correction to
the lower limit on the neutralino mass, in the case that $m_{\chi^{\pm}} 
= 85\gev$, tan$\beta = 2$ and $\mu = - 140\gev$. The plots are
obtained by varying $-500\gev \le A_t \le 500\gev$ and $50\gev \le m_0
\le 1000\gev$. The solid line in the second panel is for the fixed-point
value
$A_t = 2 m_{1/2}$. We see that the limit on $m_{\chi}$ is reduced by
about 1~GeV if we assume this value of $A_t$ and take $m_0 = 80\gev$,
corresponding to the tree-level lower limit on $m_{\chi}$.}}
{ \label{fig:scatter} }
\end{figure}

We turn first to the impact on this analysis of the radiative corrections 
discussed in the previous section.
We display in Fig.~\ref{fig:scatter} the 
radiatively-corrected
values of $m_{\chi}$ for $m_{\chi^{\pm}} = 85$~GeV, tan$\beta = 2$ 
and the representative value $\mu = -140$ GeV,
corresponding to $m_{\chi} = 40$~GeV at the tree level. The ranges 
plotted are generated by varying $A_t$ and
$m_0$ over the ranges indicated in the two 
panels. We see that their
effect is in general to reduce $m_{\chi}$ by $\sim 1$ to 
2~GeV.~\footnote{We have verified that this shift is numerically
insensitive to $\mu$ over a range larger than that of relevance to this
analysis.} If we
fix $A_t$ to the fixed-point value
discussed earlier, we obtain the solid line shown
in the second panel, and if we specify the value $m_0 \sim 80$~GeV
that corresponded to the minimum value of $m_{\chi}$ in our previous
analysis, we find a reduction in $m_{\chi}$ by $\sim 1$~GeV.
We have investigated whether the r\^oles of the LEP 172~GeV Higgs limits
and of cosmology are altered significantly when radiative corrections
are included, and found that this is not the case. We conclude that the
sensitivity of the previous LEP 172~GeV lower limit on $m_{\chi}$ was 
around the 1~GeV level.

We now turn to the implications of the LEP 183~GeV data set. Again,
for simplicity we restrict our attention to the case of universal
soft supersymmetry-breaking scalar masses, including Higgs masses,
implement the cosmological
density constraint and focus on the region tan$\beta \sim 2$ to $3$,
where we previously found the lower bound on $m_{\chi}$ under these
assumptions. The LEP lower limit on the chargino mass is now
$m_{\chi^{\pm}} \ge 91$ GeV, corresponding to a strengthening of
the lower limit on $m_{\chi}$ by $\sim 3$~GeV. Including the shift
downwards by $\sim 1$~GeV due to the radiative corrections, this
direct LEP lower limit corresponds to $m_{\chi} \ga 42$~GeV. We
have verified that the LEP Higgs searches and the cosmological
density constraint continue to exclude loopholes in this limit.

The best limit on the Higgs boson of the Standard Model is 88~GeV,
which also applies to $m_h$ for tan$\beta \la 2$
in the MSSM~\cite{latesth,contrh,LEPC,Janot}. The formulae we
use to calculate $m_h$ are believed to have an uncertainty of about
2~GeV~\cite{MSSMHiggs}, so we allow MSSM parameter sets for which they
yield
$m_h \ge 86\gev$ at low tan$\beta$. The LEP lower limit falls to about 
78~GeV for tan$\beta \ga 3$, and we allow MSSM parameter sets for
which $m_h \ge 76\gev$ for such larger values of tan$\beta$.
These Higgs limits strengthen the lower limit on $m_{\chi}$ for
tan$\beta \le 2.6$ for $\mu < 0$. 
Moreover, the strengthening of the LEP lower limit on
$m_{\chi^{\pm}}$, combined with cosmology, and including the two-loop
running of the gaugino masses as well as the one-loop corrections to
the chargino and neutralino masses, imply that $m_{\chi} \ga 
50\gev$ for tan$\beta \ge 2.8$. Only in the region
$2.6\le\tan\beta\le2.8$ for $\mu < 0$
is there a narrow cleavage where $m_{\chi}
\ga 42\gev$ is still possible.  
There is no corresponding cleavage for $\mu > 0$, where the
previous lower bound 
$m_{\chi} \ga 50\gev$ for $\tan \beta \ga 2$ remains
essentially unchanged.

As we discussed above, for sufficiently low $\tan\beta$, one cannot
satisfy simultaneously the Higgs mass constraint and the requirement
that
$t_0>12$ Gyr.  With the strengthening of the
lower limit on $m_h$ from LEP 183, we now find that tan$\beta \ga 2.0$
for $\mu < 0$ and tan$\beta \ga 1.65$ for $\mu > 0$.  Because the 
dependence of the Higgs mass on the stop masses
is only logarithmic, these
bounds on tan$\beta$ are quite insensitive to variations in $A_t$ or
$t_0$.
These strengthened results make it very difficult to reconcile our
assumptions of universal scalar masses and an interesting
cosmological relic density with the low $\tan\beta$  infra-red fixed-point solution
of the renormalization-group equations favoured by some model
builders~\cite{irqfp}.

\section{Implications for Higgsino Dark Matter}

\begin{figure}[t]
\begin{center}
\vspace*{-1.4in}
\hspace*{-0.2in}
\epsfig{file=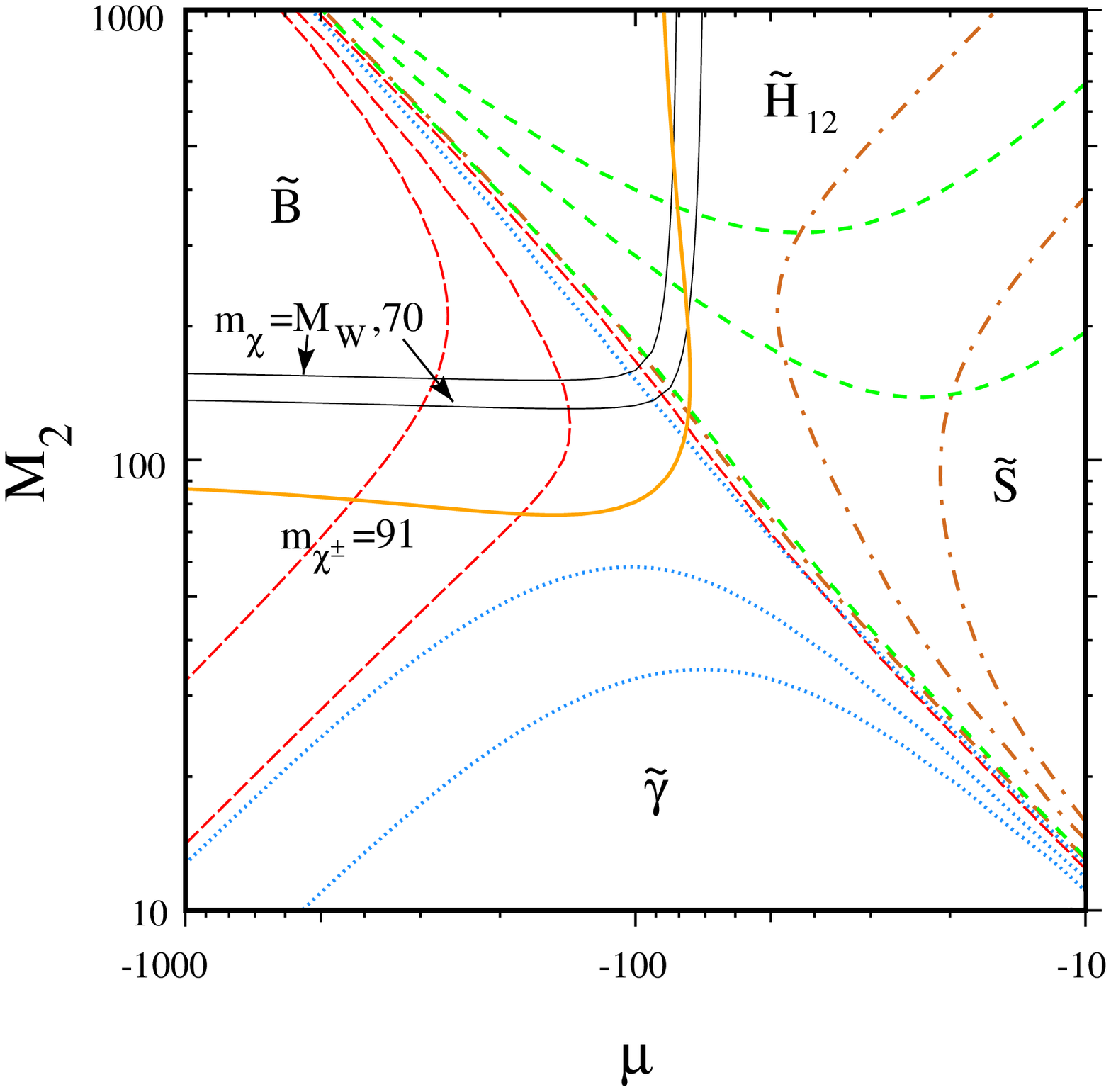,height=4.7in}
\hspace*{-0.6in}
\epsfig{file=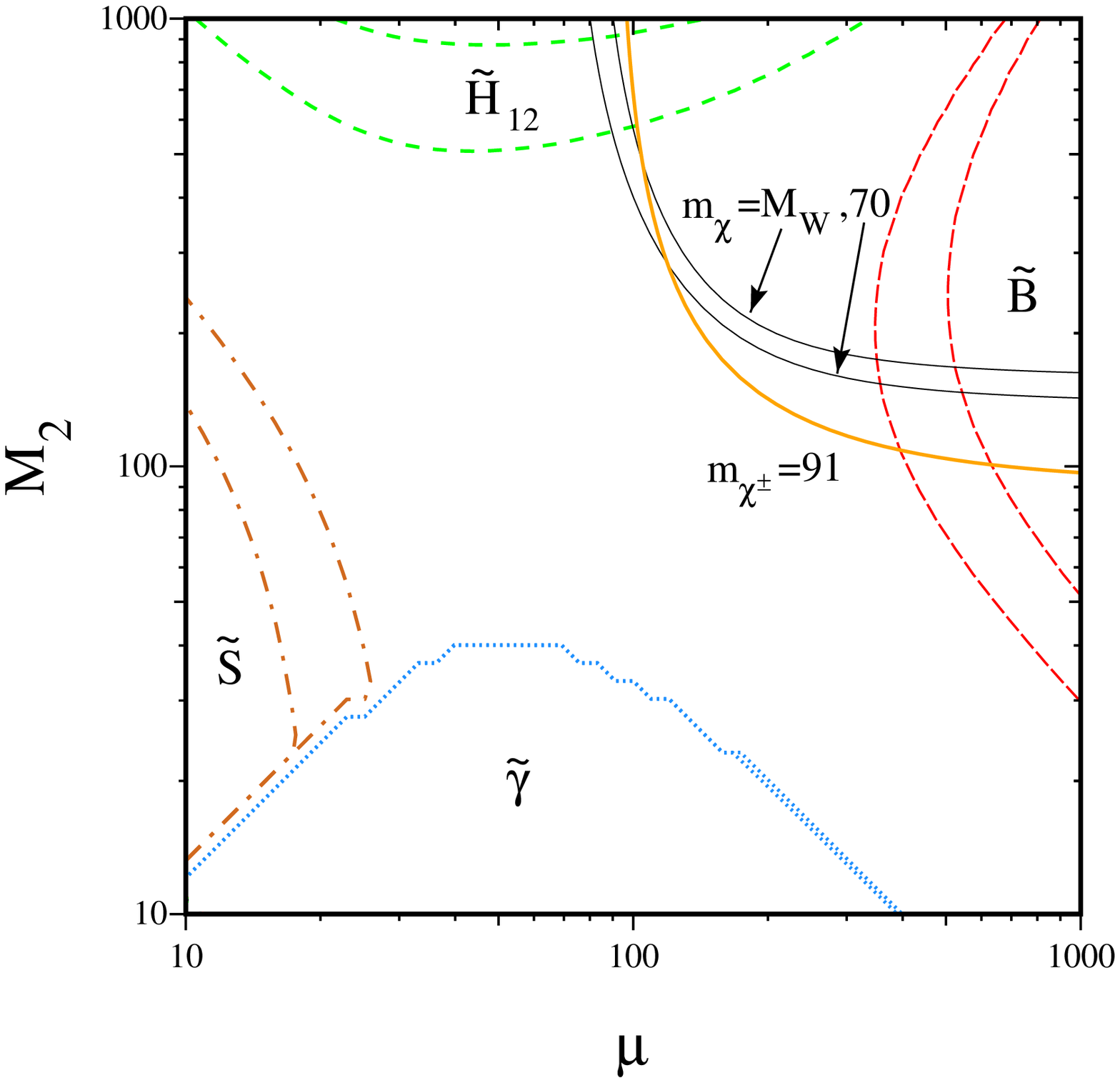,height=4.7in}
\end{center}
\vspace{-1.0in} 
\caption{{\it Contours of neutralino purity: 99\%, 97\% and (for $\mu <
0$) 75\%, and chargino and neutralino masses (solid lines). The
long-dashed lines are contours of high bino purity, the dotted lines are
contours of high photino purity, the dashed lines
are contours of high ${\tilde H}_{12}$ Higgsino purity, and the
dash-dotted lines are contours of high ${\tilde S}_0$ 
Higgsino purity. The
radiative corrections to $m_{\chi}$ are calculated for tan$\beta = 2$
and $m_0 = 100$~GeV.}}
{ \label{fig:purity} }
\end{figure}

Although constrained supersymmetric models which impose universality on
all the soft scalar masses, including the Higgs bosons, predict 
as a general feature that the
lightest neutralino is a gaugino~\cite{UHMbino}
as discussed above, it is well
known that, if the universality assumption is relaxed,
the lightest supersymmetric particle
can be a Higgsino over roughly half the parameter space as viewed 
on the $M_2$, $\mu$ plane~\cite{EHNOS}, namely when $|\mu| \la M_2$,
as seen in Fig.~\ref{fig:purity}, which displays contours of
higgsino and gaugino purities, as well as 
radiatively-corrected $m_{\chi}$ contours.
In this section we study whether it is still possible, in the light
of radiative corrections and the latest LEP data, for the
lightest neutralino to be a Higgsino, even if one relaxes
the scalar-mass-universality assumption. 

%

For this purpose, our working definition of a Higgsino is that
$p^2={\gamma_i}^2 + {\delta_i}^2 > 0.81$.  In much of the Higgsino
parameter space, it is possible to be even more definite.  For small
to moderate $M_2$, the lightest supersymmetric particle is the
Higgsino combination defined by ${\tilde S}^0 \equiv {\tilde H}_1^0
\cos \beta + {\tilde H}_2^0 \sin \beta$, i.e., $\gamma = \cos \beta$
and $\delta = \sin \beta $, with $m_{\tilde S_0} \rightarrow \mu \sin
2 \beta$~\cite{EHNOS}. Contours of ${\tilde S}^0$ purity are
displayed in Fig.~\ref{fig:purity} as
dash-dotted lines.
On the other hand, for large $M_2$, the
lightest neutralino is the state ${\tilde H}_{12} \equiv {1 \over
  \sqrt{2}} ({\tilde H}_1^0 \pm {\tilde H}_2^0)$, i.e., $\delta = \pm
\gamma = \pm 1/\sqrt{2}$ for sgn$(\mu) = \pm 1$, with $m_{\tilde
  H_{12}} \rightarrow |\mu|$~\cite{osi3}.  Contours of ${\tilde H}_{12}$
purity are shown as dashed lines in Fig.~\ref{fig:purity}.
The cross-over point where
the neutralino becomes more like $\tilde H_{12}$ than ${\tilde S}^0$
depends on $\mu$ and $\tan\beta$,  but is typically at $M_2$ of a few
hundred $\gev$~\cite{osi3}, as seen in Fig.~\ref{fig:purity}.
The ${\tilde S}^0$ interacts much
like a neutrino, i.e., it annihilates and scatters elastically via $Z$
exchange, with a coupling reduced by $\cos^2 2\beta$. On the other
hand, a ``pure" ${\tilde H}_{12}$ state never annihilates or scatters
through the $Z$, since the coupling $\gamma^2 - \delta^2 \rightarrow
0$.  Because of its similarity to a heavy neutrino, the ${\tilde S}^0$
may have an interesting relic density, with $\ohsq$ determined mainly
by its mass and $\tan \beta$. However, this same property makes it
ideal for exploration by LEP.  For example, for $\tan\beta=2$,
a ${\tilde S}^0$  state with 99\%  purity exists
only for $|\mu|<25\gev$, and from Figs.~\ref{fig:muM2} 
and \ref{fig:purity} one can see that
the region in the $\mu, M_2$ plane for such a pure state
is explored by LEP, as discussed below in more detail.
In contrast, the
${\tilde H}_{12}$ has been very difficult to find or exclude. There are
in principle two regions in the $\mu, M_2$ plane where ${\tilde H}_{12}$
can provide an interesting relic density.  For large $M_2$, they
correspond to $|\mu| < M_W$ but with $m_{\chi^{\pm}}$ above the LEP limit,
and $|\mu| \ga 1$ TeV~\cite{osi3}. 
The intermediate-mass Higgsino states are not of cosmological interest,
because of
their rapid annihilations to $W$ and $Z$ pairs. We have little to add
concerning the very  heavy ${\tilde H}_{12}$ states, but
the lighter Higgsinos lie directly in the region where the current LEP
runs are eating away at the parameter plane, as we now discuss. 

In the region where the lightest neutralino is ${\tilde H}_{12}$, this 
state is nearly degenerate with the associated chargino state,
and in some cases with the second-lightest
neutralino $\chi_2$. Therefore,
computation of the relic abundance must then take into account the
co-annihilation with these degenerate partners~\cite{co1,co2}.
The relic abundance is determined in this case by a
Boltzmann equation in which the annihilation cross-section is
generalized to include co-annihilation terms:
$\sigma_{eff} = \sum \sigma_{ij} r_i r_j$,
where $r_i$ is the relative abundance of particle $i$, 
and we include all three
nearly degenerate states.  The factors $r_i$ are in fact exponentially
sensitive to the mass differences between the states. These mass differences
are in turn sensitive to the radiative corrections to the neutralino and
chargino masses, with potentially significant
implications for the relic density, as pointed out in~\cite{DNRY}.

As a first approximation, the relic densities calculated with
and without radiative corrections are rather similar for similar
values of $\Delta M$.
Deep in the Higgsino region at very large $M_2$, 
the mass difference between the LSP and chargino
is very small, and co-annihilations are 
very important in greatly reducing the relic density to uninteresting
levels.  In this limit, the small value of $\Delta M$ also 
makes this region  inaccessible to LEP analysis.
As one decreases $M_2$, the mass difference increases and co-annihilation 
becomes less important, thus increasing the relic density
and the Higgsino's cosmological significance, while at the same time
making this region readily accessible to LEP particle searches. 

%
\begin{figure}[t]
\begin{center}
\vspace*{-1.4in}
\hspace*{-0.2in}
\epsfig{file=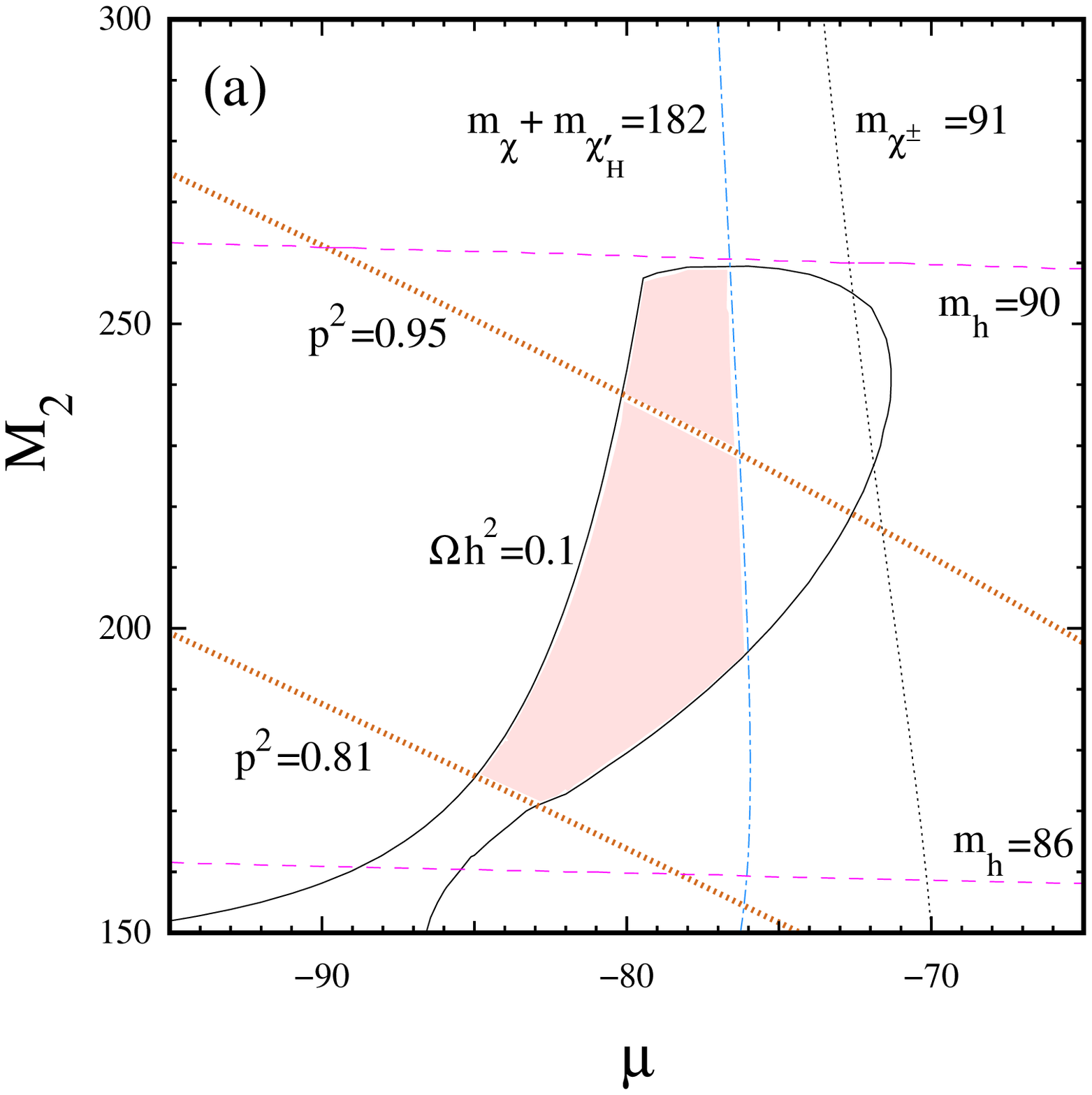,height=4.7in}
\hspace*{-0.6in}
\epsfig{file=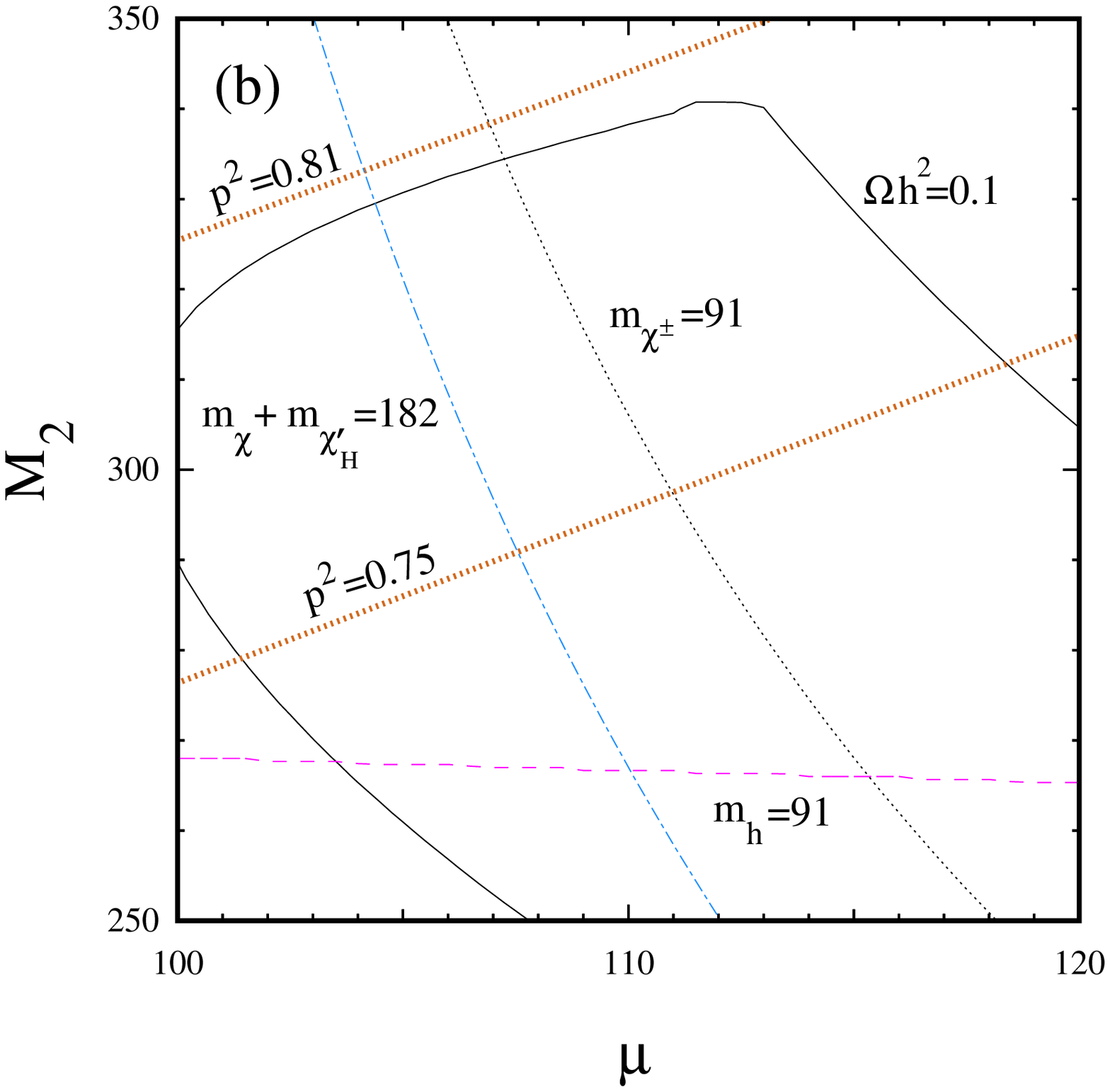,height=4.7in}
\end{center}
\vspace{-1.0in} 
\caption{{\it Survey of experimental and cosmological
constraints in the $\mu, M_2$ plane, focusing
on Higgsino dark matter for tan$\beta = 2$ and (a) $\mu < 0$ and (b) $\mu > 0$. We 
plot the radiatively-corrected contours for $m_{\chi^{\pm}} = 91\gev$,
for $m_{\chi} + m_{\chi'_H} = 182\gev$, for selected
values of $m_h$ and the Higgsino purity $p$, 
and for $\ohsq = 0.1$.  The shaded regions yield a Higgsino which satisfies
the mass and relic density constraints described in the text.}}
{ \label{fig:figsm} }
\end{figure}

In Fig.~\ref{fig:figsm} we display contours of constant
chargino mass $m_{\chi^{\pm}} = 91$~GeV, $m_{\chi} + m_{\chi'_H}
= 182$~GeV,
Higgs mass $m_h$, and Higgsino purity, along with a contour
of constant $\ohsq=0.1$, for the choice $\tb=2.0$ which is of
interest for the subsequent discussion.
Since the universal scalar mass constraint must be relaxed for
Higgsino dark matter, we have chosen
$m_A=m_0=1\tev$ so as to maximize the Higgs mass and hence
minimize the impact of this constraint.

We see that the dashed lines in Fig.~\ref{fig:figsm} 
representing the 
chargino and associated neutralino mass contours
bound one away from small $|\mu|$, whilst the Higgs mass limit 
bounds one away from small $M_2$. This is particularly
restrictive at low $\tb$, where the tree-level Higgs mass is small, and
thus where  radiative corrections to $m_h$ must be enhanced by taking
large stop masses.  
The solid contour contains the region which leads to a significant
neutralino relic density $\ohsq \ge 0.1$, and its limited range in $M_2$ is a 
result of
co-annihilations.  For larger values of $M_2$, the neutralino is a
purer Higgsino, and the masses of both the lightest chargino and the 
next-to-lightest neutralino approach the neutralino mass from
above, enhancing the effect of
co-annihilations that deplete the relic Higgsino abundance. 
For larger values of $|\mu|$, the relic density is suppressed
by annihilations into $W$ pairs. 
The hashed contours in Fig.~\ref{fig:figsm}a represent Higgsino purities.
Note the limited range of $\mu$
for which the mass and relic density constraints are satisfied.
In Fig.~\ref{fig:figsm}b we show the equivalent plot for $\mu>0$; in this case
the Higgsino purities are lower, and the entire dark matter region falls below
the purity cutoff.

%
\begin{figure}[ht]
\vspace*{-1.2in}
\begin{center}
\hspace*{-0.2in}
\epsfig{file=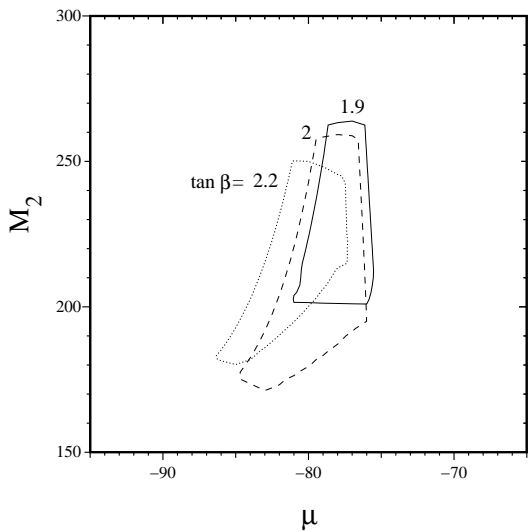,height=4.7in}
\end{center}
\vspace{-1.0in} 
 \caption{{\it The regions of the $\mu, M_2$ plane
allowed by the constraints shown in the previous figure are
shown for several different values of tan$\beta$.  There are no consistent choices of Higgsino
parameters for $\tan\beta < 1.8$ or $ > 2.5 $ for $\mu < 0$, or
for any value of $\tan\beta$ for $\mu > 0$.}}
{ \label{fig:puddles} }
\end{figure}
\begin{figure}[ht]
\begin{center}
\vspace*{-1.2in}
\hspace*{-0.2in}
\epsfig{file=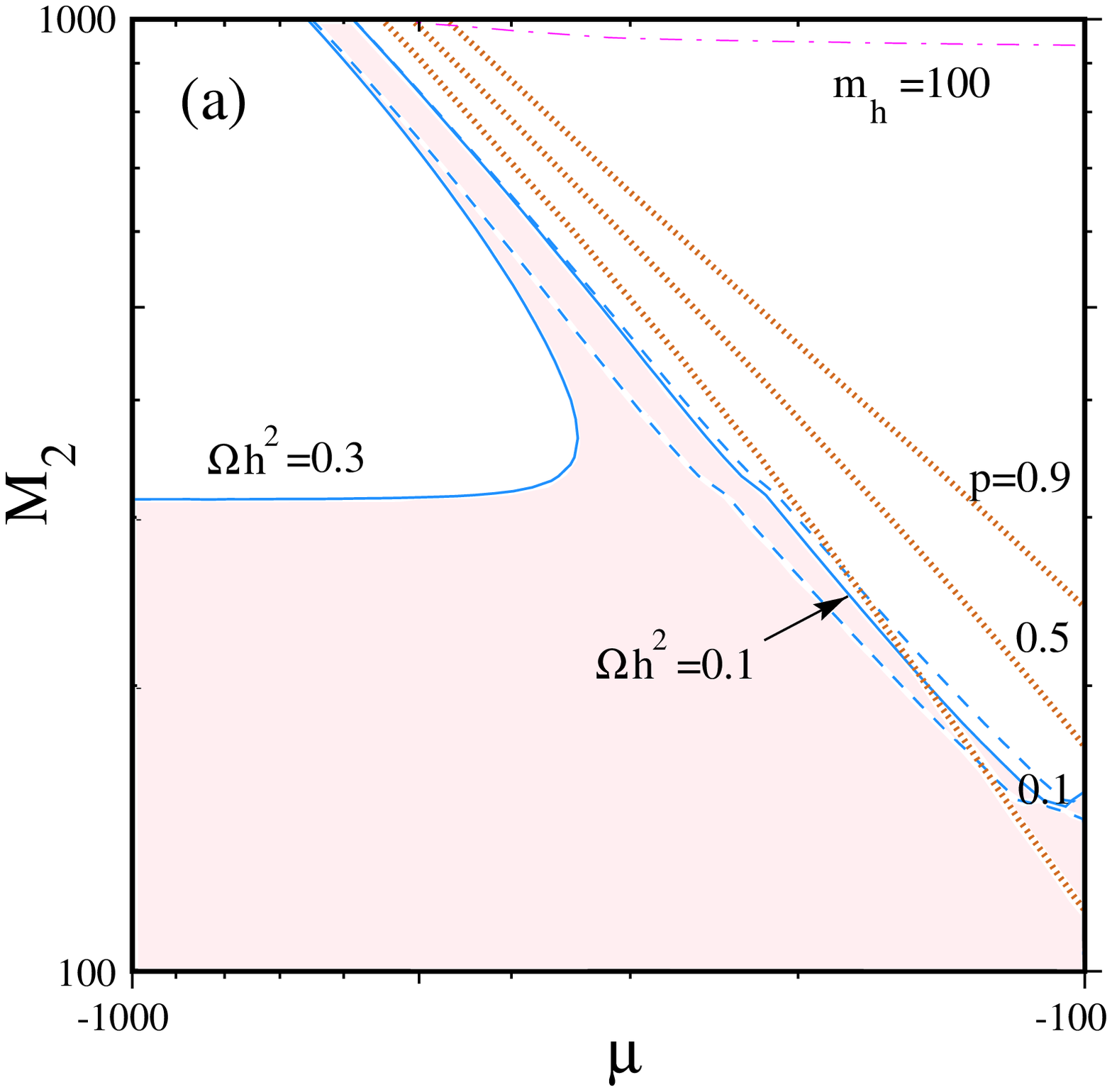,height=4.7in}
\hspace*{-0.6in}
\epsfig{file=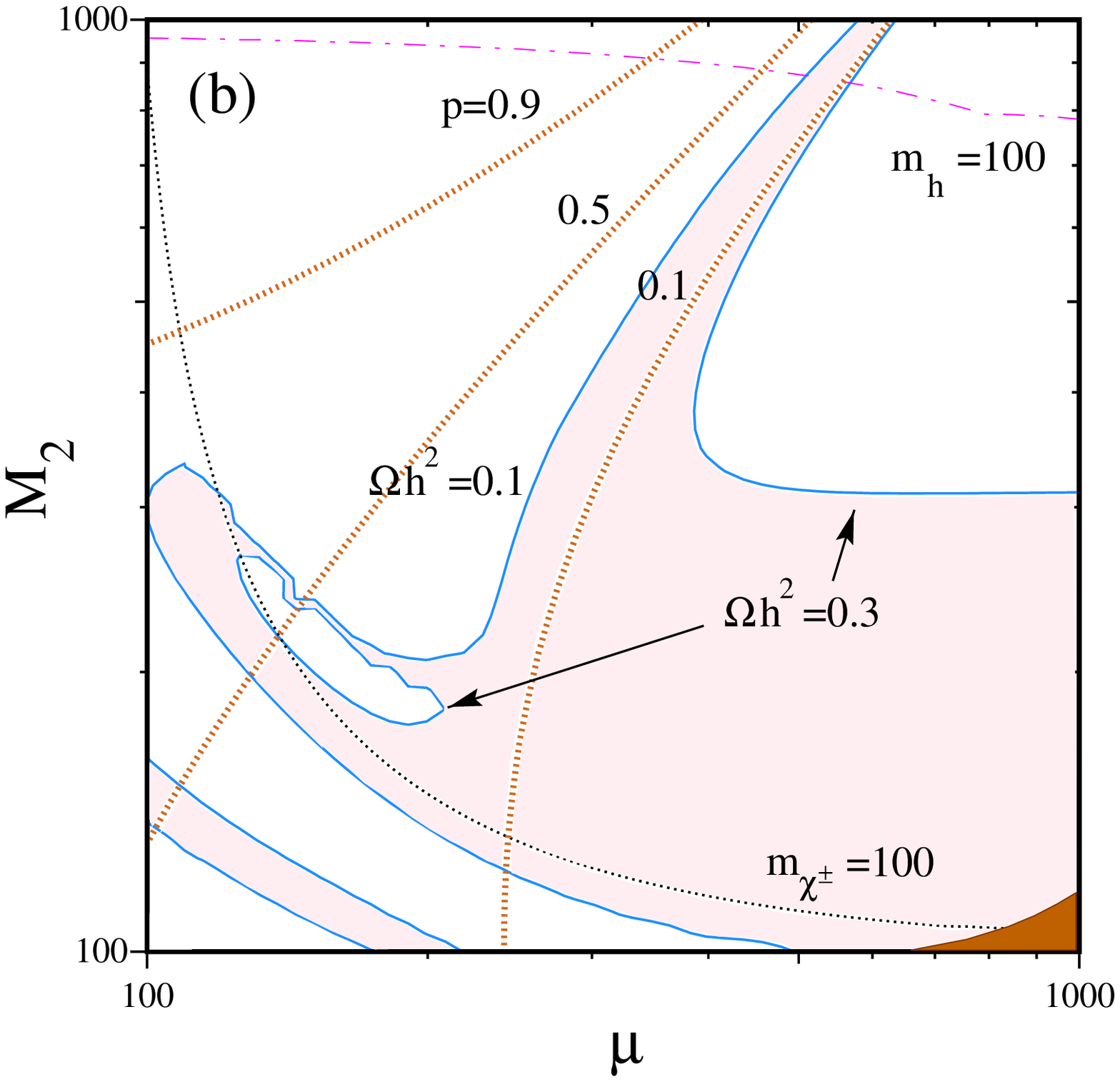,height=4.7in}
\end{center}
\vspace{-1.0in} 
\caption{{\it Larger regions of the $\mu, M_2$
plane for tan$\beta = 2$ and (a) $\mu < 0$, (b) $\mu > 0$, with
the radiatively-corrected contours for $m_{\chi^{\pm}} = 100\gev$
and $m_h = 100$ GeV,
three choices of the Higgsino
purity, and the $\ohsq = 0.1, 0.3$ contours for $m_0 = 100\gev$ and 
(in Fig.\protect\ref{fig:figbg}a)
$1000\gev$ (dashed lines). We see that the allowed regions for gaugino
dark matter
are very much larger than those for Higgsino dark matter shown in
Fig.~\protect\ref{fig:figsm}. The small dark shaded region in
Fig.~\protect\ref{fig:figbg}b yields a stop with a negative mass
squared.}}
{ \label{fig:figbg} }
\end{figure}
The combined effects of the above constraints,
corresponding to the shaded regions of Fig.~\ref{fig:figsm},
are displayed for different values of tan$\beta$ 
in Fig.~\ref{fig:puddles}. We have
again taken $m_0=m_A=1 \tev$ to minimize the effects of the Higgs and
chargino mass constraints.   We find that there
are no consistent Higgsino dark matter candidates for 
$\tan\beta \le 1.8$ or $\ge 2.5$ for $\mu < 0$, or for
any value of $\tan\beta$ for $\mu > 0$.
The Higgs mass constraint cuts off the bottom of the allowed regions at low
$\tan\beta$. When $\mu<0$ it becomes a relevant constraint for
$\tan\beta<2.0$ and is responsible for the complete disappearance of
the allowed region when $\tan\beta \le 1.8$.
Within the allowed regions displayed, the relic densities generally increase as
$|\mu|$ is increased, until the neutralino mass, whose minimum value here
is $\sim 71\gev$,  becomes greater than $m_W$, 
at which point the $W^+ W^-$ annihilation channel
opens, driving the relic $\ohsq$ below 0.1.  In any event, $\ohsq$ is
never greater than 0.12 anywhere in the allowed
regions for $\mu<0$.

If one considers neutralinos which have a lower Higgsino purity, i.e. are more mixed 
states, then additional allowed regions appear.  For example, if the purity condition is
relaxed to $p^2>0.75$,  then for $\mu>0$  there are regions of
comparable area to those in Fig.~\ref{fig:puddles} for which $\ohsq>0.1$, and these
regions extend over a range in $\tan\beta$ from 1.6 to about 25.   Similarly, 
regions considerably smaller than those in Fig.~\ref{fig:puddles} appear for $\mu<0$ at $\tan\beta>10$ and persist up to $\tan\beta$ of about
45.   At large $\tan\beta$, $A_t$ can vary somewhat from the quasi-fixed point, and 
as the Higgs mass is considerably larger than the current experimental bound, one has
the freedom to play with $m_0$ to find regions of larger $\ohsq$.  However, these regions
remain very small, regardless of $A_t$ and $m_0$.

We have taken a large value of $m_0$,
namely $1\tev$, for which the large stop masses produce large
radiative corrections to $m_h$.  On the other hand, we have chosen a
value of $m_t$ 2 $\gev$ below the current central value, and a
larger top mass would increase the radiative corrections.  
However, this is
important only for the low values of $\tan\beta$ discussed above, where
the Higgs mass constraint is important. 
On the other hand, if $\tan\beta$ is too
small, a larger top mass will lead to a Landau pole in the evolution
of $\lambda_t$.  Due to this natural ceiling on the top
mass as a function of $\tan\beta$, one cannot find any acceptable
region for $\tan\beta<1.7$ for $\mu<0$, even pushing the top Yukawa
to its perturbative maximum.  For $\mu>0$, we have already taken $m_t$
as large as possible, hence the allowed range of $\tan\beta$ cannot be
extended to lower $\tan\beta$ in this way.  

Our allowed Higgsino regions shown in Figures \ref{fig:figsm} -
\ref{fig:puddles} are somewhat smaller than those depicted in
\cite{DNRY}, where the Higgsino region 
for negative $\mu$ extends up to $M_2 \sim 400$ GeV
and down to $\mu \sim -60$ GeV for $\Omega h^2 > 0.1$.  An important
difference in our analysis is the strengthening of the LEP bounds, not
only on the
chargino mass (to about 91 GeV), but on the associated
production of Higgsinos (to $m_{\chi} + m_{\chi'_H} > 182\gev$, as seen in
Fig.~\ref{fig:figsm}).  This alone cuts out about 2/3 of the Higgsino
region of~\cite{DNRY}.  In addition, we have set the value of
$A_t$ to its infrared quasi-fixed point: $A_t \sim 2
m_{1/2} \sim 2 m_{\tilde g} / 3$~\cite{irqfp}, for the reasons discussed
above, whereas a larger
value $A_t \simeq 1$ TeV was chosen in~\cite{DNRY}. Our choice of
$A_t$ limits {\em ab initio} the size of the radiative corrections
which could remove the degeneracy between the Higgsino and chargino
states.  Thus, in our analysis the co-annihilation of these particles
remains important above $M_2 \ga 250$ GeV, in contrast to~\cite{DNRY}. 

The corrections to the
chargino and neutralino masses are positive in the regime 
studied, and also
increase with the sfermion masses, although the corrections are much
smaller than for $m_h$.  Thus the chargino and associated neutralino
production bounds tend to be more restrictive for lower $m_0$,
unless one tunes $m_{\tilde \nu}$ so that it lies close to the chargino
mass, in such a way
that the production cross section is reduced by ${\tilde \nu}$ 
exchange and
the experimental efficiency for the chargino search is reduced.
This might weaken the chargino constraint, but the associated
neutralino constraint is actually strengthened by positive
interference from ${\tilde \nu}$ exchange.
We also recall that the $\ohsq$ contours
themselves
are quite insensitive to the choice of $m_0$ in the deep Higgsino region,
so there is no great net change in the boundaries of the 
allowed Higgsino regions in Figs.~\ref{fig:figsm} and \ref{fig:puddles}.
On the other hand, as we discuss below, in regions where the gaugino
admixture is greater than in Figs.~\ref{fig:figsm} and
\ref{fig:puddles}, lowering $m_0$ decreases the relic density further
as annihilations into fermions via sfermion exchange get enhanced.  
Lastly, as discussed above,  the
value of $A_t(m_Z)$ is quite insensitive to the choice of
$A_t$~\footnote{We have
  checked what happens if we relax the gaugino mass unification
  constraint and take $M_1=M_2$~\cite{Kane}.  We find that the allowed
  regions are somewhat smaller than in the unified case for $\mu<0$
  and disappear entirely for $\mu>0$.  The latter results from a
  combination of larger neutralino masses - so that $m_\chi> m_W$ in a
  greater region, and closer $\chi^{\pm} - \chi$ mass degeneracy -
  leading to a higher co-annihilation rate.}.

The areas $\cal{A}$ of the allowed Higgsino regions are never very large:
calculating them using the logarithmic measure $(d\mu / \mu)(d M_2 / 
M_2)$, we find that their maxima are reached for 
tan$\beta \sim 2$ for both $\mu < 0$ and $\mu > 0$, with ${\cal
A}_{max} = 0.020, 0.006$, respectively. 
Furthermore, the residual region for $\mu < 0$
could be explored by a search sensitive either to $m_h \le 100\gev$ or 
$m_{\chi} + m_{\chi'_H}\le 200$~GeV,
as anticipated if the LEP energy is eventually increased to
200~GeV, and the $\mu > 0$ region could be explored by a search for
$m_{\chi^{\pm}} \le 100\gev$~\footnote{A fraction of the (very small)
large $\tan\beta$ regions which open up for $\mu<0$ when the purity
constraint is relaxed would not be probed by the LEP run at 200~GeV}.
To put these numbers in perspective,
one might wish to compare them with the logarithmic area of the full
$\mu, M_2$ plane for values of these parameters between 100 and
1000~GeV,  namely ${\cal A} \sim 5$. 
Alternatively, one might wish to compare with the logarithmic area
of  the parts of this plane where gaugino dark matter is still
viable. This can be visualized in Fig.~\ref{fig:figbg}, where we
display the regions where $0.1 \le
\ohsq \le 0.3$ for $m_0 = 100\gev$ (solid contours) and $m_0=1000\gev$
(dashed contours in Fig.~\ref{fig:figbg}a). 
Here we choose
$m_A=3000\gev$ to avoid the annihilation pole associated with the
heavy Higgs, but the other parameter values
are as in Fig.~\ref{fig:figsm}. 
In the region with high gaugino
purity and low Higgsino content, annihilation is predominantly through
sfermion exchange into fermion
pairs.  For $m_0=1000\gev$, annihilation through this channel is
effectively shut off, leading to the very small allowed region
demarcated by the dashed contours in Fig.~\ref{fig:figsm}a.    For
$m_0=100\gev$,  the cosmologically preferred region is significant.
It is clear that the gaugino area is
much larger than the remnant Higgsino area, indicating that this is a
much more generic possibility. Indeed, the logarithmic area for gaugino
dark matter is ${\cal A} \ga 2$ for both $\mu$ positive and negative,
some two orders of magnitude larger than the residual Higgsino regions. 

\section{Conclusions}

The main purpose of this paper has been to include radiative corrections 
to chargino and neutralino masses for the first time in a
phenomenological analysis of the LEP data. This analysis has been
based on an implementation~\cite{FG} of the analytical results published
previously in one of the codes developed for LEP data
analysis. This implementation
makes it possible for radiative corrections to be taken into account
in experimental analyses.
In our first application of the radiatively-corrected code~\cite{FG},
we have found that radiative corrections
generically reduce the upper limit on the lightest neutralino mass by
about 1~GeV compared to what would be inferred from a tree-level analysis.

We have also updated our previous analysis of constraints from LEP and
cosmology to include preliminary results from the latest LEP run at 183~GeV.
Assuming scalar-mass universality, the previous lower limit on $m_{\chi}$
is now increased to about 42~GeV,
and the previous lower bounds on tan$\beta$ are increased to about 2.0
for $\mu < 0$ and 1.65 for $\mu > 0$, making life difficult for infra-red 
fixed-point models.

We have also studied in detail the viability of Higgsino dark matter in
the light of the latest LEP constraints, relaxing the
scalar-mass-universality assumption. The radiative 
corrections play an essential r\^ole, as do the LEP constraints
on charginos, associated neutralino production and Higgs bosons. We
find that $m_\chi \ga 71$~GeV in the remaining Higgsino region,
that for $\mu < 0$ there is only a very limited range of tan$\beta$ over
which Higgsino dark matter is viable at all, and that the area of the
the $\mu, M_2$ plane within which it is viable is always very
restricted. This is particularly evident when the Higgsino region is
compared with the region of the $\mu, M_2$ plane over which gaugino 
dark matter is viable.

Finally, we review briefly the prospectives for further exploration
of neutralino dark matter with higher-energy runs of LEP. We have already
observed that prospective improvements in the sensitivity to MSSM Higgs
bosons, charginos and other supersymmetric particles should enable the
complete investigation of neutralino dark matter up to $\sim 50$~GeV,
if universal input scalar masses and an interesting cosmological relic
density are assumed. Additionally, the analysis of this paper indicates
that
the future higher-energy LEP runs should also provide closure on the
possibility of Higgsino dark matter, by telling us whether the
lighter chargino or the lightest MSSM Higgs boson weighs less than
100 GeV.

\vskip 0.5in
\vbox{
\noindent{ {\bf Acknowledgments} } \\
\noindent  
This work was supported in part by DOE grant DE--FG02--94ER--40823.  The
work of T.F. was supported in part by DOE   
grant DE--FG02--95ER--40896 and in part by the University of Wisconsin  
Research Committee with funds granted by the Wisconsin Alumni Research  
Foundation.}

\end{document}